\documentclass[3p,times,procedia]{elsarticle}
\usepackage{nupha_ecrc}

\volume{00}

\firstpage{1}

\journalname{Nuclear Physics A}

\runauth{}

\jid{nupha}

\jnltitlelogo{Nuclear Physics A}

\usepackage{amssymb}

\usepackage[figuresright]{rotating}

\begin{document}

\begin{frontmatter}



\dochead{}

\title{Critical point search from an extended parameter space of lattice QCD at finite temperature and density}


\author[label0,label1]{Shinji Ejiri}
\fntext[label0]{Presented by S. Ejiri}
\author[label2]{Ryo Iwami}
\author[label3,label4]{Norikazu Yamada}

\address[label1]{Department of Physics, Niigata University, Niigata 950-2181, Japan}
\address[label2]{Graduate School of Science and Technology, Niigata University, Niigata 950-2181, Japan}
\address[label3]{KEK Theory Center, Institute of Particle and Nuclear Studies, High Energy Accelerator Research Organization (KEK), Tsukuba 305-0801, Japan}
\address[label4]{School of High Energy Accelerator Science, SOKENDAI (The Graduate University for Advanced Studies), Tsukuba 305-0801, Japan}

\begin{abstract}
Aiming to understand the phase structure of lattice QCD at nonzero temperature and density, we study the phase transitions of QCD in an extended parameter space, where the number of flavor and quark masses are considered as parameters.
Performing simulations of 2 flavor QCD and using the reweighting method, 
we investigate $(2+N_f)$ flavor QCD at finite density, where two light flavors and $N_f$ massive flavors exist. 
Calculating probability distribution functions, we determine the critical surface terminating first order phase transitions in the parameter space of the light quark mass, the heavy quark mass and the chemical potential. 
Through the study of the many flavor system, we discuss the phase structure of QCD at finite density.
\end{abstract}


\end{frontmatter}


\section{Introduction}
\label{sec:intro}

The study of the critical point at finite density is one of the most interesting topics in QCD.
The chiral phase transition at the physical quark masses is considered to be crossover at low density, but it is expected to change into a first order transition at the critical density.
However, numerical simulations of lattice QCD at high density are difficult due to the sign problem. 
Using the property that the nature of the transition changes also with the quark mass, 
it will be important to investigate the boundary of the first order transition region at low density in the quark mass parameter space of (2+1) flavor QCD. 

\begin{figure}[t]
\begin{center}
\vspace*{-2mm}
\centerline{
\includegraphics[width=45mm,clip]{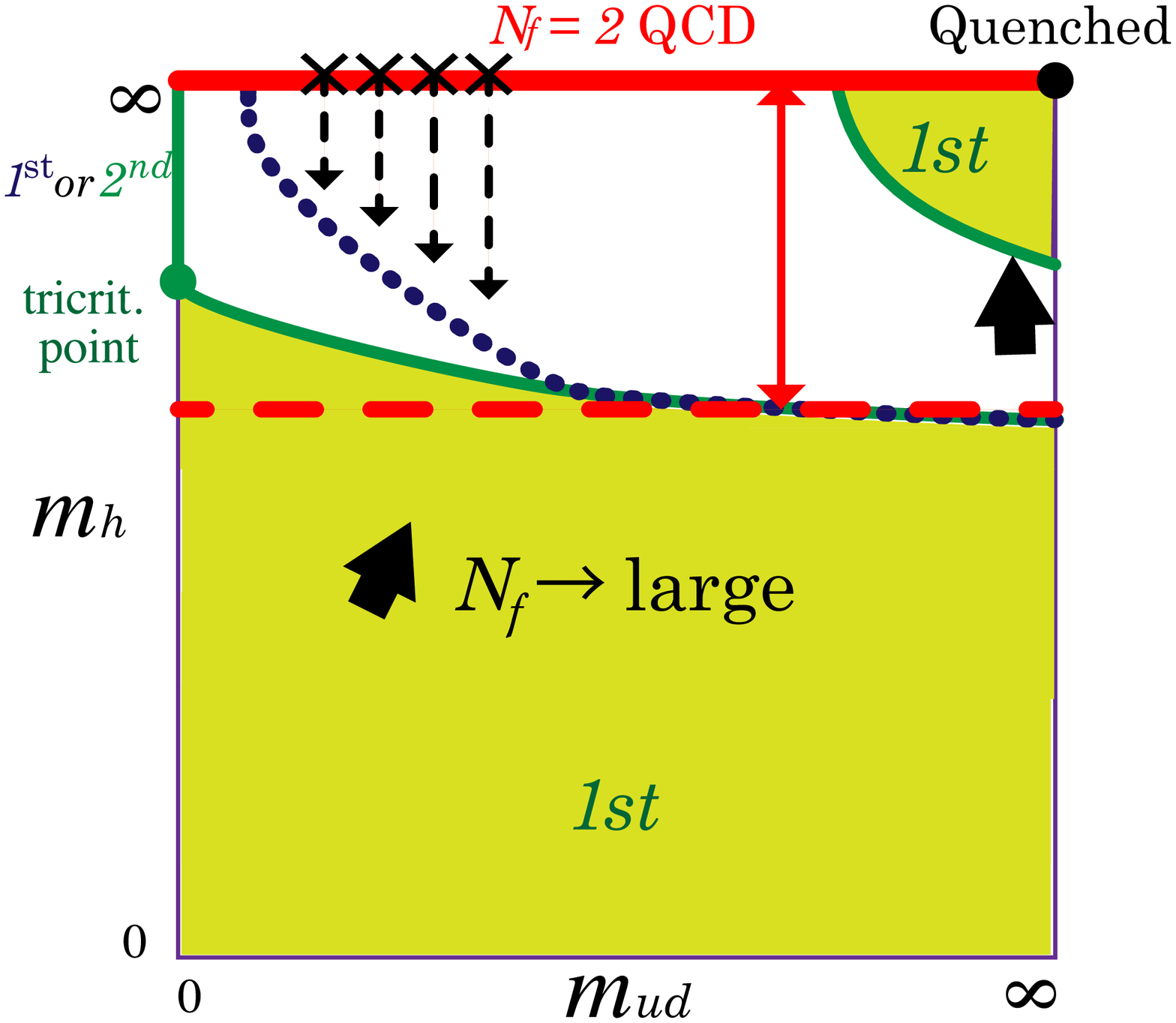}
\includegraphics[width=60mm,clip]{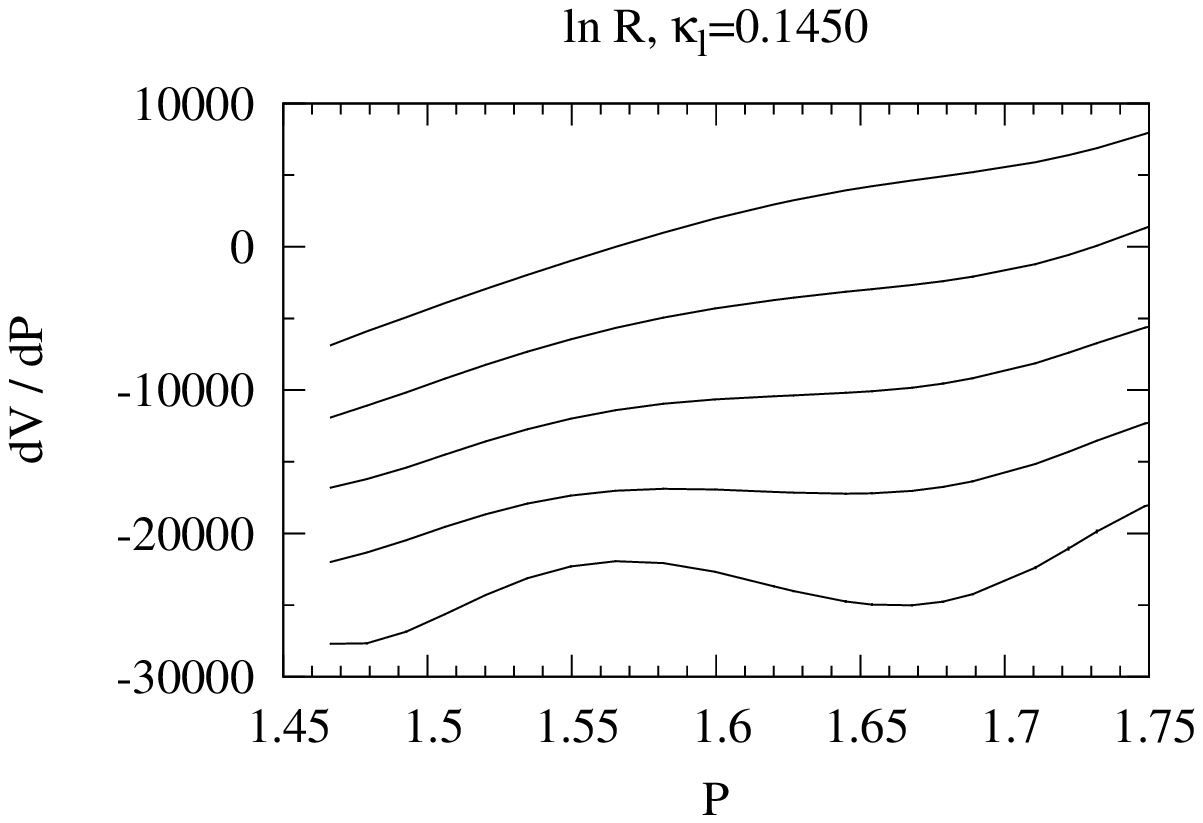}
}
\vspace{-2mm}
\caption{Left: The quark mass dependence of the nature of phase transition (Columbia plot) for $(2+N_f)$ flavor QCD.
Right: The slope of the effective potential $V_{\rm eff}(P)$ for $h=0.0$ - $0.4$ from 
top to bottom.}
\label{fig1}
\end{center}
\end{figure}

In this paper, we study a system in which two light quarks and $N_f$ heavy quarks exist, extending the parameter space.
In Ref.~\cite{yamada13}, the critical heavy quark mass is found to increase as $N_f$ increases and the endpoint of the first order transition can be investigated in the heavy quark region, 
where the hopping parameter expansion works well, for large $N_f$.
We illustrate the quark mass dependence of the nature of phase transitions in Fig.~\ref{fig1} (left) for (2+$N_f$) flavor QCD at zero density, where $m_{ud}$ and $m_h$ are the masses of light flavors and $N_f$ flavors.
The phase transition is of first order in the yellow regions and the green curves are the second order critical curves separating the first order and crossover regions. 

We focus on two topics. 
One is the finite temperature phase transition of the massless 2 flavor QCD \cite{yamada15}.
To understand the nature of the massless limit is a long standing problem.
Investigating the light quark mass dependence of the boundary of the first order region, one may find whether the massless limit of 2 flavor QCD, i.e. top left corner of Fig.~\ref{fig1} (left), is of first order or second order. 
If the boundary runs like the dotted curve in Fig.~\ref{fig1} (left), the transition is of first order in the massless limit.
The other is the chemical potential dependence \cite{iwami15}.
We will show that the critical $m_h$ increases as the chemical potential $\mu_h$ increases. 
This suggests that the critical curve can be investigated in the heavy quark region even for small $N_f$ when $\mu_h$ is large.
Through the study of $(2+N_f)$ flavor QCD, we discuss the QCD critical point at finite density.

\section{Histogram method}
\label{sec:method}

We study the system where 
two light flavors and $N_f$ heavy flavors exist.
The hopping parameter and the quark chemical potential are $\kappa_l$, $\mu_l$ for the light flavor and $\kappa_h$, $\mu_h$ for the heavy flavor, respectively. 
The hopping parameter is in inverse proportion to the quark mass when the mass is heavy.
To investigate the nature of phase transitions, we consider the probability distribution function of the average plaquette, 
$
 w(P;\beta, \kappa_{l}, \mu_l, \kappa_h, \mu_h) 
 = \int \mathcal{D} U \delta \left( P - \hat{P} \right) 
e^{6 \beta N_{\rm site} \hat{P}} \left( \det M (\kappa_l,\mu_l) \right)^{2} 
\left( \det M (\kappa_h,\mu_h) \right)^{N_f}, 
$
where $M$ is the quark matrix, 
$N_{\rm site} \equiv N^{3}_s \times N_t$ is the number of lattice sites,
and $\beta = 6/ g^{2}_{0}$ is the coupling constant. 
$\hat{P}$ is defined from the gauge action $S_g$ as $\hat{P} = -S_{g} /(6N_{\rm site} \beta)$, which is a linear combination of Wilson loops and 
is called the generalized plaquette.
$\delta (P - \hat{P})$ is the delta function, which constrains the operator $\hat{P}$ to $P$.
We use the delta function approximated by
$\delta(x) \approx 1/(\Delta \sqrt{\pi})$ $\exp[-(x/\Delta)^2]$, 
where the parameter $\Delta$ must be small.
For convenience, we define the effective potential as 
$V_{\rm eff} (P) \equiv - \ln w(P)$.
It is rewritten as 
\begin{eqnarray}
 V_{\rm eff}(P; \beta, \kappa_{l}, \mu_l, \kappa_h, \mu_h) 
 = V_{0} (P;\beta_{0}, \kappa_l) - \ln R(P; \beta, \beta_0, \kappa_l, \mu_l, \kappa_h, \mu_h) 
\end{eqnarray}
with the potential of 2 flavor QCD at $\mu_l =0$, $V_0 (P;\beta_0, \kappa_l)$, and
the reweighting factor 
\begin{eqnarray}
 \ln R (P) 
 &=& 6 (\beta - \beta_0) N_{\rm site} P + \ln \left\langle \left( \frac{ \det M (\kappa_l, \mu_l) }{ \det M (\kappa_l, 0) } \right)^{2} \left(    \frac{ \det M (\kappa_h, \mu_h) }{ \det M (0, 0) } \right)^{N_f} \right\rangle_{(P; {\rm fixed})} ,
\label{eq:lnr}
\end{eqnarray}
where 
$\left\langle \cdots \right\rangle_{(P; {\rm fixed})}$ 
means the ensemble average over 2 flavor configurations when the plaquette is fixed to be $P$.
We evaluate the quark determinant of $N_f$ flavors in $R(P)$ with the leading order of the hopping parameter expansion for the standard Wilson action \cite{whot13}. 
\begin{eqnarray}
\ln \left[ \frac{ \det M(\kappa_h, \mu_h) }{ \det M(0,0) } \right]
&=& 288 N_{\rm site} \kappa^{4}_{h} \hat{W}_P + 12 N^{3}_{s} ( 2 \kappa_{h} )^{N_t} \left( \cosh (\mu_h /T) \hat{\Omega}_{R} + i \sinh (\mu_h/T) \hat{\Omega}_{I} \right) + \cdots , \ \ \
\label{eq:lnrhpe}
\end{eqnarray}
where
$\hat{W}_P$ is the $(1 \times 1)$ Wilson loop,
$\hat{\Omega}_{R}$ and $\hat{\Omega}_{I}$ are the real and imaginary parts of the Polyakov loop.
This approximation is valid when $\kappa_h$ is small or the quark is heavy.
Note that no expansion in terms of $\mu_h/T$ is performed. 
On the other hand, the quark determinant of light flavors is computed by a Taylor expansion in terms of $\mu_l /T$ \cite{BS02, ejiri08}, assuming $\mu_l /T$ is small.
Because the sign problem is serious for the calculation of $R(P)$ at high density, 
the method discussed in Ref.~\cite{ejiri08, whot09} is used to avoid the sign problem.

We then find the critical $\kappa_h$, at which the first order transition terminates.
At a first order transition point, $V_{\rm eff}$ shows 
a double-well shape as a function of $P$, 
and equivalently the curvature of the potential $d^2 V_{\rm eff}/dP^2$ 
is negative around the center of the double-well potential.
Moreover, the curvature $d^2 V_{\rm eff}/dP^2$ is independent of $\beta$, since $\beta$ appears only in the linear term of $P$ in the right hand side of Eq.~(\ref{eq:lnr}).
While $\beta$ must be adjusted to the first order transition point to observe the double-well potential, the fine tuning is not necessary if we investigate the curvature \cite{ejiri08}.
Hence, we investigate the curvature of the potential and find the endpoint of the first order phase transition by calculating $V_0$ and $\ln R$ at simulation points. 

\begin{figure}[t]
\begin{center}
\vspace{-8mm}
\centerline{
\hspace*{-8mm}
\includegraphics[width=60mm,clip]{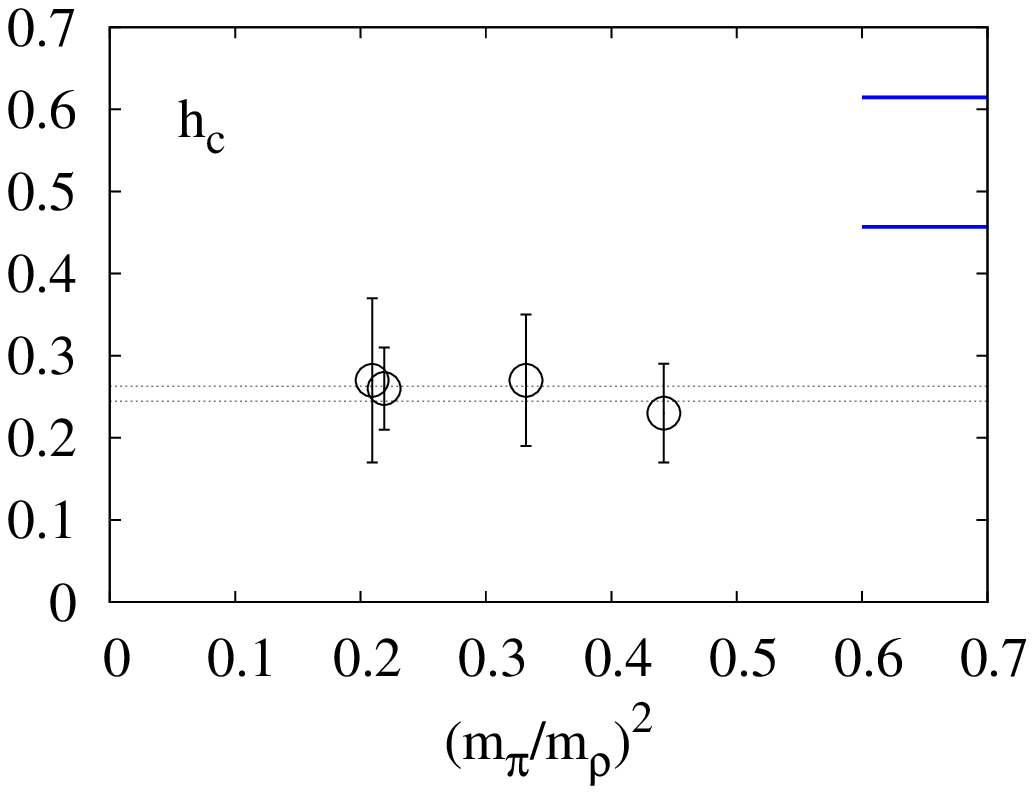}
\hspace*{-9mm}
\includegraphics[width=50mm,clip]{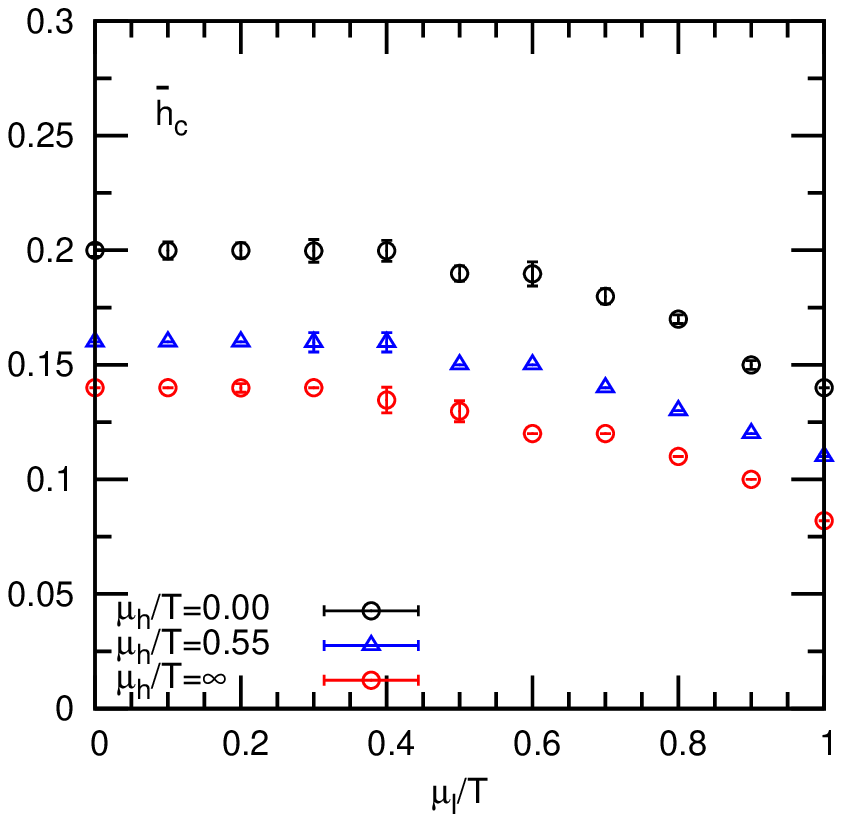}
\hspace*{-5mm}
\includegraphics[width=55mm,clip]{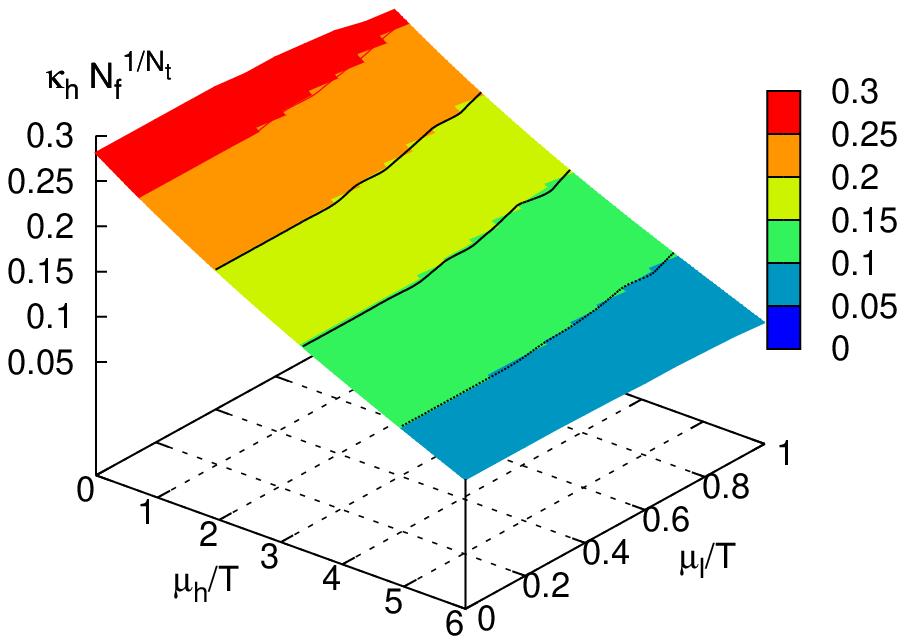}}
\vspace{-4mm}
\caption{Left: The critical value of $h$ at $\mu_l =\mu_h =0$.
Middle: The preliminary result of $\bar{h}_c = h_c \cosh (\mu_h /T)$ at $\kappa_l = 0.1475$.
``$\mu_h/T=\infty$'' means $\tanh (\mu_h/T)=1$.
Right: The critical surface in the $(\kappa^{c}_{h} N_f^{1/N_t}, \mu_l /T, \mu_h /T)$ 
space at $\kappa_l = 0.1475$.}
\label{fig2}
\end{center}
\end{figure}

\section{Boundary of the first order region}
\label{sec:result}

We perform simulations of QCD with degenerate 2 flavor $O(a)$-improved Wilson quark and RG-improved Iwasaki gauge actions at zero density.
The lattice size $N_{\rm site}$ is $16^{3} \times 4$.
We adopt four hopping parameters of light quarks $\kappa_l= 0.1450$ to $0.1505$ and  
25 to 32 $\beta$ values at each $\kappa_l$, which cover the pseudo-critical $\beta$. 
10,000 to 40,000 trajectories are accumulated at each simulation point.
These values of $\kappa_l$ correspond to the mass ratio of pseudoscalar and vector mesons  
$m_{\pi} / m_{\rho}= 0.665$ to $0.458$.
The details of the simulation parameter are shown in Ref.~\cite{yamada15}.
We use all configurations for zero density and the finite density analysis is carried out with 500 configurations taken every 10 trajectories at $\kappa_l = 0.1450$ and $0.1475$ among the configurations. 
The light quark determinant is evaluated up to $O(\mu_l^2)$.

\subsection*{Light quark mass dependence}

Since $\beta$ must be adjusted to be the critical value to study the phase transition, the reweighting factor $\ln R$ is controlled only by $h \equiv 2 N_f (2 \kappa_h)^{N_t}$ for $N_t=4$ at zero density as seen in Eq.~(\ref{eq:lnrhpe}).
We evaluate the critical $h$, $h_c$, instead of $\kappa_h$.
The right panel of Fig.~\ref{fig1} shows the first derivative $dV_{\rm eff} / dP$ at $\kappa_l=0.145$ and $\mu_l=\mu_h=0$ for $h=0.0, 0.1, 0.2, 0.3$ and $0.4$ from top to bottom as an example.
$dV_{\rm eff} / dP$ is monotonically increasing at $h=0$ while it is an ``S''-shaped function at $h=0.4$. 
This means that the shape of the effective potential changes into double-well type at $h_c$.
The critical values of $h$ are determined by the results of $d^2 V_{\rm eff} / dP^{2}$, where the error is dominated by the systematic uncertainties associated with the fitting procedure for the calculation of derivatives of $V_{\rm eff}$.
We thus evaluate $h_c$ changing the fitting procedure, the width of the approximate delta function etc. to estimate the systematic error. 
The details are given in Ref.~\cite{yamada15}.
The light quark mass dependence of $h_c$ is plotted as a function of
$\left( m_\pi/m_\rho \right)^2$ in Fig.~\ref{fig2} (left). 
$h_c$ determined from the direct $2+N_f$ flavor simulation with $\kappa_l =0$ and $N_f =50$ is shown with a sizable uncertainty by two blue lines at the top right corner.
While the fact that $h_c$ in $\kappa_l =0$ is clearly larger than those around
$0.145 \le \kappa_l \le 0.1505$ may indicate that $h_c$ decreases towards the
chiral limit, $h_c$ stays constant in that range of $\kappa_l$ and 
the critical heavy quark mass seems to remain finite in the chiral limit.
The nonzero value of $h_c$ in the chiral limit suggests that the transition of massless 2 flavor QCD is of second order.

\subsection*{Chemical potential dependence}

The first order transition also arises at large $\mu_l$ and $\mu_h$ even when $h$ is small.
As in Eq.~(\ref{eq:lnrhpe}), the heavy quark determinant is controlled by 
$\bar{h} \equiv h \cosh (\mu_h /T) =2N_f (2\kappa_h)^{N_t} \cosh (\mu_h /T)$
and $\tanh (\mu_h/T)$. 
We determine the critical value of $\bar{h}$ for each $\tanh (\mu_h/T)$.
The Wilson loop term in Eq.~(\ref{eq:lnrhpe}) can be absorbed into the gauge action, thus we omit this term.
The preliminary result of $\bar{h}_c$ is shown in Fig.~\ref{fig2} (middle) as functions of $\mu_l /T$ for $\kappa_l=0.1475$ \cite{iwami15}. 
Here, the systematic error which arises in this analysis has not been estimated yet. The error bar represents the statistical error only.

In this analysis, we have used the approximation by a Taylor expansion up to $O(\mu_l^2)$, thus the analysis is valid only in the region where $\mu_l/T$ is small.
On the other hand, the $\mu_h$ dependence is easy to investigate, since $| \tanh (\mu_h/T) |$ is smaller than one. 
We find that the first order region becomes wider as $\mu_l$ and $\mu_h$ increase in the mass parameter space.
The qualitative behavior is consistent with the previous results obtained by an improved staggered fermion action \cite{yamada13}.
However, the quantitative difference from the staggered fermion is not small. 
Although the systematic error is not yet estimated, the discretization error of our result by the $N_t=4$ lattice may be large. 
The critical value $\bar{h}_c$ is not very sensitive to $\tanh(\mu_h/T)$.
The difference between the results of $\tanh(\mu_h/T)=0$ and $1$ is only about $30\%$.
This situation is similar to the case of the critical surface in the heavy quark region 
(top right corner of Fig.~\ref{fig1} (left)) \cite{whot13}.

The critical $\kappa^{c}_{h} N_f^{1/N_t}$ decreases exponentially with $\mu_h$ for large $\mu_h$ when $\tanh(\mu_h/T)$ dependence of $h_c$ is small.
We plot the critical value of $\kappa^{c}_{h} N_f^{1/N_t}$ in Fig.~\ref{fig2} (right) as a function of $\mu_l/T$ and $\mu_h/T$ translating Fig.~\ref{fig2} (middle) \cite{iwami15}. 
Because the approximation of $O((\mu_l/T)^2)$ is used, $\mu_l/T$ must be small in this analysis.
Whereas, the critical points are determined for $0 \leq \tanh(\mu_h/T) \leq 1$, i.e. 
$0 \leq \mu_h/T \leq \infty$.
In this study, we have assumed $\kappa_h$ to be small to apply the hopping parameter expansion 
and this condition is always valid for large $N_f$ even if the critical $h$ is large.
However, for large $\mu_h$, $N_f$ is not needed to be large because 
$\kappa^{c}_{h} N_f^{1/N_t}$ is no longer large. 
This means that our analysis can be applied for small $N_f$ such as $(2+1)$ flavor QCD when the density of strange quark is very large.
This suggests that the extrapolation from the high strange quark density region may be a possible approach to the determination of the QCD critical point.

\section{Conclusions and outlook}
\label{sec:conc}

We studied the finite temperature and density phase transition of QCD with two light flavors and $N_f$ massive flavors.
Through the shape of the distribution functions, we determined the critical surface separating the first order transition and crossover regions.
The light quark mass dependence suggests that phase transition of the massless 2 flavor QCD is of second order, and the critical heavy quark mass becomes larger as the chemical potential increases
at finite density.

Another interesting extension of the parameters is QCD with a complex chemical potential. 
The effective potential also becomes first-order-transition-like when the chemical potential is a complex number \cite{yoneyama09,yoneyama13,yoneyama15}.
This property make a singularity, which is called ``Lee-Yang zero''. 
The nature of phase transitions can be investigated from the singularities in the complex parameter plane. 
The study in the complex parameter space may provide important information about the QCD phase transition.





\bibliographystyle{elsarticle-num}
\bibliography{<your-bib-database>}



\end{document}